\documentclass[prl,twocolumn,showpacs]{revtex4}
\usepackage{epsf}
\usepackage{epsfig}
\begin{document}
\title{Stable Equilibrium Based on L\'evy Statistics: Stochastic Collision Models Approach} 
\author{Eli Barkai}
\affiliation{Dept. of Chemistry and Biochemistry, Notre Dame University, Notre Dame, IN 46556.}
\date{\today}

\begin{abstract}

We investigate equilibrium properties of two very different
stochastic collision
models: (i) the Rayleigh particle and (ii) the driven Maxwell gas.
For both models the  equilibrium velocity distribution 
is a L\'evy distribution, the Maxwell distribution being a special case.
We show how these models are related to fractional kinetic
equations. Our work demonstrates that a stable power-law equilibrium,
which is independent of details of the underlying models,
is a natural generalization of Maxwell's velocity distribution.

\end{abstract}


\maketitle

 There is a strong analogy between the Gaussian central limit 
theorem (GCLT) and the relaxation to thermal equilibrium of the Boltzmann
equation (\cite{BC} and Ref. therein).
 However the GCLT is non-unique, which may imply that
standard thermal equilibrium is non-unique. The L\'evy
central limit theorem (LCLT) considers the case of summation
of independent identically distributed random variables with
an {\em infinite} variance \cite{Feller}. Hence following  Montroll and
Shlesinger  it is  natural
to  ask if generalized equilibrium
concepts based on LCLT are
meaningful 
\cite{Montroll}.
Here we start answering this question using 
two very different types of collision models which
still reveal the same type of equilibrium.
We note that L\'evy statistics has many physical applications \cite{review},
however its possible relation to generalized forms
of equilibrium statistical mechanics
is an
open field of research.
Recently, Bobylev and Cercignani \cite{BC}, investigated a non-linear
Boltzmann equation with an infinite velocity variance showing
that the solution exists, and obtaining certain bounds on
it. 
In \cite{BC} the possibility
of a relation between solutions
of the Boltzmann equation \cite{Ernst}
and LCLT was briefly pointed out.

 Our goal in this  Letter is to show that a new equilibrium
concept naturally emerges from old stochastic collision
models. Our models demonstrate that:  
(i) L\'evy velocity distributions serve as
the natural generalization of the Maxwell velocity distribution,
(ii) 
generalized power law equilibrium can be derived from
kinetic models, {\em there is no need to postulate a specific form
of power law equilibrium}, and
(iii) the L\'evy equilibrium obtained here possesses a certain domain
of attraction, is unique, and does not depend on certain
details of the underlying models.

%
%

{\bf Model 1}  We consider a one dimensional tracer particle with the mass $M$
coupled with gas particles of mass $m$. The tracer particle velocity
is $V_M$. At random times the tracer particle collides
with gas particles whose velocity is denoted with $\tilde{v}_m$.
Collisions are elastic hence from conservation of momentum and energy
$V_M ^{+} = \xi_1 V_M ^{-} + \xi_2 \tilde{v}_m$,
where 
$\xi_1 = { 1 - \epsilon \over 1 + \epsilon}$, $\xi_2 = { 2 \epsilon \over 1 + \epsilon} \ $,
$\epsilon \equiv  m/M$ is the mass ratio and $V_M ^{+}$
$(V_M ^{-})$
is the velocity of the tracer particle after (before) a collision event. 
The duration of the collision events is much shorter than
any other time scale in the problem.
The collisions occur at a uniform rate $R$.
 The probability density function (PDF)
of the gas particle
velocity is $f(\tilde{v}_m)$. This PDF does not change during the collision
process, indicating that re-collisions of the gas 
particles and the tracer particle are neglected.

 Many works considered this type of model,
imposing the condition that the gas particles are distributed
according to Maxwell's law, i.e $f(\tilde{v}_m)$
is Maxwellian. 
Since we are now investigating possible
generalizations of Maxwell's law we change this strategy and
assume that $f(\tilde{v}_m)$ is non-Maxwellian. The goal is
to see when and how the tracer particle reaches a universal
equilibrium, which does not depend on the detailed shape
of 
$f(\tilde{v}_m)$. 

 We now consider the equation of motion for the tracer particle
velocity PDF $W(V_M , t)$ with initial conditions concentrated on
$V_M(0)$. Standard kinetic considerations yield
the linear Boltzmann equation
\begin{widetext}
\begin{equation}
 {\dot W}\left(V_M ,t \right)  = 
 - R W\left( V_M, T \right)
 + R \int_{-\infty}^{\infty} {\rm d} V_M ^{-} \int_{-\infty}^{\infty} {\rm d} \tilde{v}_m  
W \left(V_M ^{-}, t\right)f\left( \tilde{v}_m \right) \times 
\delta\left( V_M - \xi_1 V_M ^{-} - \xi_2 \tilde{v}_m \right),
\label{eq03}
\end{equation}
\end{widetext}
where the delta function gives the constrain on energy and
momentum conservation in collision events. Us-usual the first (second) term in
Eq. (\ref{eq03}) describes a  tracer particle leaving (entering)
the velocity point $V_M$ at time $t$.
Eq. (\ref{eq03}) contains convolution integrals in velocity
space hence we consider now its Fourier transform (FT). Let
$\bar{W}(k,t)$
 be the FT of the velocity PDF $W(V_M,t)$. 
%
%
Using Eq. (\ref{eq03}), the equation of motion for 
$\bar{W} \left( k , t \right)$  is a finite difference
non-local equation
\begin{equation}
\dot{\bar{W}}(k,t)  = - R \bar{W}(k,t) +
R \bar{W} \left( k \xi_1 , t \right) \bar{f} \left(k \xi_2  \right),
\label{eq05}
\end{equation}
where $\bar{f} \left( k \right) $ is the FT of
$f(\tilde{v}_m)$. 
The solution 
of the equation of motion Eq. (\ref{eq05}) is obtained by iterations
\begin{equation}
\bar{W}\left( k, t \right) = 
\sum_{n=0}^{\infty} { \left( R t \right)^n \exp\left( - R t \right)  \over n!} e^{ i k V_M(0) \xi_1^n } \Pi_{i=1}^n \bar{f}\left( k \xi_1^{n-i} \xi_2 \right).
\label{eq06}
\end{equation}
This solution has a simple interpretation.
The probability that the tracer particle has collided $n$ times with
the gas particles is given according to the Poisson law
$ P_n(t)= { \left( R t  \right)^n \over n!} \exp\left( - R t \right)$.
Let $W_n(V_M)$ be the PDF of the tracer particle conditioned that
the particle experiences $n$ collision events. It can be shown
that the FT of
$W_n(V_M)$ is
$\bar{W}_n(k) = 
 e^{ i k V_M(0) \xi_1^n } \Pi_{i=1}^n \bar{f}\left( k \xi_1^{n-i} \xi_2 \right).
$
Thus Eq. (\ref{eq06}) is a sum over the probability of having $n$
collision events in time interval $(0,t)$ times the FT
of the velocity PDF after exactly $n$ collision event.

 In the long time limit  
$\bar{W}_{eq}(k) \equiv \lim_{t \to \infty} \bar{W}(k, t )$
 an equilibrium is  obtained from Eq. (\ref{eq06}). We notice that when
$R t \to \infty$,  $P_n(t) = (R t)^n \exp(- R t)/n!$ is peaked
in the vicinity of $\langle n \rangle = R t$ hence it 
is easy to see that
\begin{equation}
\bar{W}_{eq}\left( k\right) = \lim_{n\to \infty}\Pi_{i=1}^n \bar{f}\left( k \xi_1^{n-i} \xi_2 \right).
\label{eq13a}
\end{equation}
In what follows  we  investigate properties of
this equilibrium. We note that similar equilibrium can be obtained also
if the collision process is not described by the Poisson law,
any $P_n(t)$ which is peaked on $n \to \infty$ when $t \to \infty$,
with (nearly) zero support for finite values of $n$,
will exhibit this behavior.
 
 We will consider the Rayleigh weak collision limit $\epsilon \to 0$. 
This limit is important since number of collisions needed for
the tracer particle to reach an equilibrium is very 
large. Hence in this case we may expect the emergence of
a general equilibrium concept which is not sensitive to the
precise details of the velocity  PDF $f(\tilde{v}_m )$ 
of the gas particles. In this limit we may also
expect that in a statistical sense $\tilde{v}_m \ll V_M$, hence
the assumption of a uniform collision rate is reasonable in this
limit.  

 We assume that statistical properties of gas particles
velocities can be characterized with an energy scale $T$.
Since $T$, $m$ and $\tilde{v}_m$ are the only variables
describing the gas particle we have
\begin{equation}
f\left( \tilde{v}_m \right) = {1 \over \sqrt{ T / m} } q\left( {\tilde{v}_m \over \sqrt{ T/ m } } \right).
\label{eqSca01}
\end{equation}
We also assume that $f(\tilde{v}_m )$ is an even function,
as expected from symmetry. 
The dimensionless function
$q(x)\ge 0$ satisfies a normalization condition
$\int_{-\infty}^{\infty} q(x) {\rm d} x = 1$,
otherwise it is rather general.
The scaling assumption made in Eq. (\ref{eqSca01}) is very natural,
since the total energy of gas particles is nearly conserved.

 We first consider the case where moments of $f\left( \tilde{v}_m \right)$
are finite. The second moment of the gas particle velocity is
$\langle \tilde{v}_m ^2 \rangle = { T \over m} \int_{-\infty}^{\infty} x^2 q(x) {\rm d} x$.
Without loss of generality we set 
$\int_{-\infty}^{\infty} x^2 q(x) {\rm d} x=1$.
The scaling behavior Eq. 
(\ref{eqSca01})
yields
$\langle \tilde{v}_m ^{2 n} \rangle = \left( { T \over m} \right)^n q_{2 n}$,
where the moments of $q(x)$ are defined according to
$q_{2n} = \int_{-\infty}^{\infty} x^{2 n} q(x) {\rm d} x$.
Thus the small $k$ expansion of the gas particle characteristic function is
\begin{equation}
  \bar{f}\left( k \right) = 
 1 - { T k^2 \over 2 m } + q_4 \left( { T \over m} \right)^2 { k^4 \over 4!} + O(k^6).
\label{eqSca06}
\end{equation}
Inserting Eq. (\ref{eqSca06}) in Eq.  
(\ref{eq13a})
we obtain
\begin{equation}
 \ln \left[ \bar{W}_{eq}\left(k\right) \right] = 
- { T \over 2 m} g_2 \left( \epsilon \right) k^2 +
{q_4 -3 \over 4!} \left( {T \over m} \right)^2 g_4 \left( \epsilon \right) k^4 + O(k^6),
\label{eqSca07}
\end{equation}
where $g_n(\epsilon) = (2 \epsilon)^n/ [ (1+\epsilon)^n - (1 - \epsilon)^n]$.
The interesting thing to notice is that in the limit $\epsilon \to 0$,
the second term on the right hand
side of Eq. (\ref{eqSca07}) is zero, thus $q_4$ is an irrelevant parameter
in the problem.
In similar way one can show that all terms
in the expansion containing $q_{2 n}$ with $n> 1$  
vanish in the Rayleigh limit $\epsilon \to 0$.
Thus  using Eq. 
(\ref{eqSca07})
\begin{equation}
 \lim_{\epsilon \to 0} \ln \left[ \bar{W}_{eq}\left(k\right) \right] = 
- { T \over 2 M} k^2. 
\label{eqSca07a}
\end{equation}
From Eq. (\ref{eqSca07a}) it is easy
to see that  the Maxwell velocity PDF for the tracer particle
$M$ is obtained. 
Thus as expected Maxwell's equilibrium is stable in the sense that
for a large class of gas particle velocity PDFs Maxwell equilibrium
is obtained. 

\begin{figure}[htb]
\epsfxsize=20\baselineskip
\centerline{\vbox{
        \epsfig{file=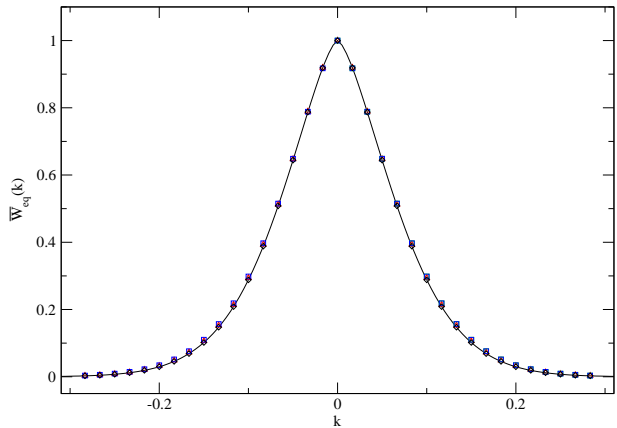, width=0.85\linewidth, angle=-90}  }}
\caption {
We show the FT of the equilibrium velocity distribution
of the tracer particle. 
Numerically exact solutions
of the problem are obtained using three long tailed 
gas particle
velocity PDFs defined in text: 
case 1 squares,  case 2
triangles,   
and case 3 diamonds.
The tracer particle equilibrium
is well approximated by the L\'evy distribution 
the solid curve;
$\bar{W}_{eq}(k)\sim\exp\left(-2.211|{k\over \epsilon^{1/6} }|^{3/2} \right)$. 
For the numerical results we used $T=4.555$, $\epsilon=1e-5$ and $M=1$.
}
\label{fig6}
\end{figure}


 Now we assume that $f\left( \tilde{v}_m \right)$ has a power law behavior,
i.e., $q(x)\propto |x|^{-(1 + \alpha)}$ when $|x| \to \infty$
and $0<\alpha<2$. For this case the gas particle characteristic function
is
 $$ \bar{f}\left( k \right) = $$
\begin{equation}
1 - { q_{\alpha} \over \Gamma\left( 1 + \alpha \right) } \left( { T \over m} \right)^{\alpha/2} |k|^{\alpha} 
+ { q_{\beta} \over \Gamma\left( 1 + \beta \right) } \left( { T \over m} \right)^{\beta/2} |k|^{\beta} 
+ o(|k|^{\beta})
\label{eqSca07b}
\end{equation}
where $\alpha< \beta \le 2 \alpha$.
$q_{\alpha}$ and $q_{\beta}$ are dimensionless numbers
which depend of-course on $q(x)$. Without loss of generality we may
set $q_{\alpha} = 1$.
In Eq. (\ref{eqSca07b}) we have used the assumption that $f(\tilde{v}_m)$
is even.

Inserting Eq. (\ref{eqSca07b}) in Eq.  
(\ref{eq13a})
we obtain a small $k$ expansion
$ \ln \left[ \bar{W}_{eq}\left( k \right) \right]$.
Taking the limit $\epsilon \to 0$ one can show that the 
terms containing $q_{\beta}$ are much smaller than the leading term
$\ln \left[ \bar{W}_{eq}\left( k \right) \right]\propto - |k|^{\alpha}$. 
Thus
$q_{\beta}$, and in a similar way higher order coefficients,
become the irrelevant parameters of the problem.   
 Thus we find that the tracer particle 
equilibrium characteristic function is    
\begin{equation}
\bar{W}_{eq} \left( k \right) \sim \exp\left[ - { 2^{\alpha-1} \over \alpha \Gamma\left( 1 + \alpha\right) } \left( { T \over M} \right)^{\alpha/2} {|k|^{\alpha} \over \epsilon^{1 - \alpha/2} } \right],
\label{eqSca100}
\end{equation} 
thus a L\'evy velocity distribution for the tracer particle
is obtained.
For $\alpha \ne 2$
the equilibrium Eq. (\ref{eqSca100}) depends
on $\epsilon$, while for the Maxwell's case $\alpha=2$, 
the equilibrium is independent of the coupling constant $\epsilon$.
Eq. (\ref{eqSca100}) implies that variance of the velocity diverges
when $\alpha<2$.
For the non-Maxwellian case 
the velocity distribution is characterized by the scale
$(T/M)^{\alpha/2}$ which determines the width of the velocity
distribution.


The asymptotic behavior Eq. (\ref{eqSca100}) is now 
demonstrated using  numerical examples.
We consider three types of gas particle velocity PDFs,
for large values of
$|v_m| \to \infty$ these PDFs exhibit 
  $f\left( \tilde{v}_m \right) \propto |\tilde{v}_m|^{-5/2}$,
namely $\alpha = 3/ 2$.
 Case 1
$f( \tilde{v}_m) =  N_1/  ( 1 + 3^{2/3}\sqrt{  m / (4 T)}  |\tilde{v}_m|)^{5/2}$, 
 Case 2, 
$f( \tilde{v}_m ) =  N_2 / [ 1 +  m \tilde{v}_m ^2 /(0.439 T) ]^{5/4} $,
where $N_1$ and $N_2$ are normalization constants.  
Case 3, the gas particle velocity PDF is a 
L\'evy PDF with index $3/2$ whose FT
is
$\bar{f}\left( k \right) = \exp\left[ - \left( { T \over m} \right)^{3/4} {  |k|^{3/2} \over \Gamma\left( 5/2 \right) } \right]$.

 According to our theory these power law velocity PDFs,
yield a L\'evy equilibrium for the tracer particle when the mass ratio
becomes small,
Eq. (\ref{eqSca100}). In Fig. \ref{fig6} we show numerically
exact solution of the problem for cases (1-3). These
solutions, obtained using Eq.  
(\ref{eq13a})
for finite values of $\epsilon$,
show a good agreement between numerical results
and the asymptotic theory. The L\'evy equilibrium for the
tracer particle is not sensitive to precise shape of
the velocity distribution of the gas particle, and hence like the
Maxwell distribution is stable. 

 We now consider a Fokker--Planck equation which
describes the evolution of the tracer particle PDF
$W(V_m,t)$
towards the L\'evy equilibrium Eq. 
(\ref{eqSca100}). The equation is of fractional
order and 
is obtained using a small $\epsilon$ expansion
of  Eq. 
(\ref{eq03}) (details to be published) 
\begin{equation}
{ \partial W(V_M,t) \over \partial  t} \simeq  {\bar{D} \over \epsilon^{1 - \alpha/2}}  {\partial^{\alpha}  W\left( V_M , t \right) \over \partial |V_M|^{\alpha}}  + \gamma  {\partial  \over \partial V_M } \left[ V_M W(V_M , t )\right] .
\label{FFeq07z}
\end{equation}
In Eq. (\ref{FFeq07z}) 
the Riesz fractional derivative was used \cite{review},
and the dissipation term is  
$\gamma = 2 \epsilon R$. A generalized Einstein
relation 
\begin{equation}
\bar{D} = {2^{\alpha - 1} \over \Gamma\left(1 + \alpha\right) } \left( { T \over M} \right)^{\alpha/2} \gamma,
\label{FFeq08}
\end{equation}
yields the relation between the transport coefficients $\bar{D}$ and $\gamma$.
When $\alpha=2$ the Einstein relation is recovered.
Note that 
\cite{Metz,Jep,Chechkin,Zannette}
investigated related fractional processes based on a stochastic approach
(e.g. Langevin Eqs. with L\'evy noise).
In those investigations dissipation 
and fluctuations 
were treated as though they are independent, hence the equilibrium
obtained there differs from ours.

{\bf Model 2}  The question remaining is wether L\'evy equilibrium
a general feature, which might be obtained
from other collision models. Specifically, it is
 interesting to see if 
L\'evy equilibrium is compatible with a non--linear
Boltzmann equation approach. Since one may suspect that the L\'evy behavior
obtained so far is limited to linear Boltzmann models.
For this aim we investigated the one dimensional driven inelastic
Maxwell model (DIMM). This model was investigated extensively
in recent years in the context of inelastic gases assuming finite
variance boundary conditions (see details below) 
\cite{Santos,BenNaimRev}.
Our goal is to investigate DIMM in the quasi elastic limit 
showing that L\'evy statistics describes the equilibrium,
the Maxwell-Gauss distribution is recovered in the proper limit.


First consider the inelastic Maxwell model in the {\em absence}
of external driving forces
$\dot{W}(V, t )  = I(V, W)$,
where the non-linear collision integral is
\begin{equation}
I(V, W) \equiv - W(V,t) + { 1 \over p} \int_{-\infty}^{\infty}  W(u,t)
W\left( { v - q u \over p}, t \right) {\rm d} u.
\label{eq02}
\end{equation}
In Eq. (\ref{eq02})
$p=(r + 1)/2$ 
and $q=1-p$
where $r$ is the restitution coefficient $0< r \le 1$.
The kinetic scheme describes a situation where momentum is conserved
during collision events, while energy is conserved only
when $r=1$. If $r<1$ the steady state solution of Eq. (\ref{eq02})
is $W_{ss}(V)= \delta(V)$, reflecting the loss of energy during collision
events. 
Note that for elastic collisions $r=1$, any initial
velocity distribution is a steady state solution. 
This is expected (and not informative) since
two identical 1D elastic particles,
exchange their velocities in collision events.

Let $\bar{W}(k,t)$ be the FT of $W(V,t)$.
The boundary conditions we will consider are
\begin{equation}
\bar{W}\left( k , t \right) \sim 1 - { \langle |V|^{\alpha} \rangle |k|^\alpha \over \Gamma\left( 1 + \alpha\right) },
\label{eq07}
\end{equation}
for small $k$.
Since $W(V,t)$ is a non-negative PDF we have $0<\alpha\le 2$.
Using the Boltzmann equation (\ref{eq02}) it is easy to
show that in the elastic limit $r=1$, 
${\partial \langle |V|^{\alpha} \rangle\over \partial t}=0$. 
For the standard case $\alpha=2$ considered in \cite{Santos},
Eq. (\ref{eq07}) simply reflects energy conservation,
i.e. 
$\langle V^2 \rangle$ is a constant of motion.
For $\alpha<2$, 
$\langle |V|^{\alpha} \rangle$ describes the width of the 
probability packet, which for elastic collision is a conserved
quantity.

 As mentioned, when $r<1$ the inelastic collisions
will shrink any initial probability packet to 
be concentrated on  $V=0$.
Similar to previous work \cite{Santos}
an {\em infinitesimal}
 heating term is added to the equation of
motion, which compensates the energy loss.
We will consider the boundary conditions
described in Eq. (\ref{eq07}), while \cite{Santos}
considered the Gaussian case $\alpha=2$.
To obtain behavior compatible with Eq. 
(\ref{eq07}) we consider the fractional DIMM
\begin{equation}
{\partial W(V, t ) \over \partial t} - D_{\alpha} {\partial^\alpha W(V,t) \over \partial |V|^{\alpha} } = I(V, W).
\label{eq08}
\end{equation}
It is more convenient to consider this fractional
equation in Fourier space, this yields
the non-linear and non-local equation
\begin{equation}
\dot{\bar{W}} \left( k , t \right)  + \left( 1 + D_{\alpha} |k|^{\alpha} \right) \bar{W}\left( k , t \right) = \bar{W} \left( p k , t \right) \bar{W} \left( q k , t \right).
\label{eq09}
\end{equation}
For our aim this equation gives the definition of the fractional derivative
in Eq. (\ref{eq08}).
Our aim is to investigate the steady state solution
of this  equation in the quasi elastic limit when
$D_{\alpha} \to 0$ and $r \to 1$. This limit is taken in such 
a way that the boundary condition Eq. (\ref{eq07}) is satisfied.

Using the condition
${\partial \langle |V|^{\alpha} \rangle \over \partial t} = 0$,
and  Eqs.
(\ref{eq07},
\ref{eq09})
we obtain
\begin{equation}
D_{\alpha} = {\langle |V|^{\alpha} \rangle \over \Gamma\left( 1 + \alpha \right) } \left( 1 - p^{\alpha} - q^{\alpha} \right).
\label{eq12}
\end{equation}
Without loss of generality we may set now $\langle |V|^{\alpha} \rangle =1$.
For $\alpha= 1$  we have $D_{\alpha}=0$, while
for $0<\alpha<1$  $D_{\alpha}$ obtains negative values.
The case $\alpha=1$ marks the transition between a
finite $(\alpha>1)$ and infinite $(\alpha<1)$ first
order moment of velocity $\int_{-\infty}^{\infty}| V |W(V,t) {\rm d} V$.
For $\alpha<1$ no steady state is obtained, since
the dissipation due to collisions is not strong enough to compensate
the heating. 
Our results in what follows are
restricted to $1<\alpha\le 2$.

 Steady state solution of Eq. 
(\ref{eq09})
 satisfy 
\begin{equation}
 \left( 1 + D_{\alpha} |k|^{\alpha} \right) \bar{W}_{ss}\left( k \right) = \bar{W}_{ss} \left( p k \right) \bar{W}_{ss} \left( q k \right).
\label{eq13}
\end{equation}
An iteration method is used to obtain the solution,
let $\psi(k) \equiv \ln\left[ \bar{W}_{ss} \left( k \right) \right]$,
and using Eq. (\ref{eq13}) we have
\begin{equation}
\psi \left(k\right) = \psi\left( p k \right) + \psi\left( q k \right) - \ln\left[ 1 + D_{\alpha} |k|^{\alpha} \right] .
\label{eq14}
\end{equation}
The boundary condition Eq.
(\ref{eq07})  yields
$\psi(k) \sim - |k|^{\alpha}/ \Gamma\left( 1 + \alpha \right)$.
The solution of Eq. (\ref{eq14}) is obtained using the
iteration rule
\begin{equation}
\psi_{n+1}(k) = \psi_n\left( p k \right) + \psi_n \left( q k \right) - \ln \left[ 1 + D_{\alpha} |k|^{\alpha} \right],
\label{eq15}
\end{equation}
where $\lim_{n \to \infty} \psi_n(k) = \psi(k)$
and the `initial condition' is
$\psi_0(k) = - \ln\left[ 1 + D_{\alpha} |k|^{\alpha} \right]$.
Using these rules and some algebra involving series expansions, 
we find
%
%
%
\begin{equation}
\psi(k) = - \sum_{n=1}^{\infty} { \left( - 1 \right)^{n + 1} |k|^{\alpha n} D_{\alpha} ^n \over n \left( 1 - q^{\alpha n } - p^{\alpha n } \right)},
\label{eq18}
\end{equation}
where the condition $1< \alpha \le 2$ was used.
Inserting Eq. (\ref{eq12})
in Eq. (\ref{eq18}) we obtain
\begin{equation}
\psi(k) = - { |k|^{\alpha} \over \Gamma\left(1 + \alpha\right) } +
{\alpha(1 - r) \over 8} { |k|^{2 \alpha } \over \left[ \Gamma\left(1 + \alpha\right) \right]^2}+ O(|k|^{3 \alpha})
\label{eq18a}
\end{equation}
The interesting thing to notice, is that the second term on the right
hand side of Eq. (\ref{eq18a})
vanishes when the elastic limit
$r \to 1$ is considered.
Inserting Eq. (\ref{eq12})
in Eq. (\ref{eq18}) one can show that  in the elastic limit
$\lim_{r \to 1} \psi(k) = - |k|^{\alpha}/  \Gamma\left( 1 + \alpha \right)$.
Hence the steady state characteristic function is
a stretched exponential
\begin{equation}
\lim_{r \to 1} \bar{W}_{ss} ( k ) = \exp\left[ - {|k|^{\alpha} \over \Gamma\left( 1 + \alpha \right)} \right],
\end{equation}
the inverse FT of this equation yields the
symmetric stable L\'evy density \cite{Feller}.
It is rewarding to find that Maxwell and L\'evy equilibrium
are obtained only in the elastic limit, thus
conservation of energy in the collision events is
related to the L\'evy--Maxwell behavior. Far from this limit
results not directly related to the Gauss--L\'evy
central limit theorem
are obtained, Eq. (\ref{eq18}).

 To conclude, we  demonstrated the relation between equilibrium
properties of very different types of
collision models and L\'evy statistics.
Thus stable behavior transcends
details of individual models and hence I suspect can be found
in other types of collision models.
To support the idea that L\'evy velocity distribution
might be found in other models, we note the interesting
work of Min et al \cite{Min} who used numerical simulations
of a long-range interacting vortex model, and showed that 
distribution of velocity fields are L\'evy stable.

\end{document}